# Investigation of Vortex Structures in Gas-Discharge Nonneutral Electron Plasma: II. Vortex Formation, Evolution and Dynamics

N. A. Kervalishvili

Iv. Javakhishvili Tbilisi State University, E. Andronikashvili Institute of Physics
Tbilisi 0177, Georgia.  <n_kerv@yahoo.com>

**Abstract.** The results of experimental investigations of inhomogeneities of gas-discharge nonneutral electron plasma obtained by using the nonperturbing experimental methods [N.A. Kervalishvili, arXiv:1502.02516 [physics.plasm-ph] (2015)] have been presented. Inhomogeneities are the dense solitary vortex structures stretched along the magnetic field, the lifetime of which is much greater than the time of electron-neutral collisions. The processes of formation, evolution and dynamics of vortex structures were studied. The periodic sequence of these processes is described for different geometries of discharge device.

## I. Introduction

Nonneutral plasmas are the ensemble of charged particles of only (or predominantly) one sign, owing to which they have the properties strongly differing them from quasineutral plasma. First of all, among the peculiarities of nonneutral plasma are the strong internal electric fields and the absence of the charges of opposite sign that makes difficult the shielding of electrostatic perturbations caused by the external electric fields. Due to this fact, one of the most widespread and universal methods of local diagnostics of low-density plasma - the Langmuir probe turned to be inapplicable for nonneutral plasmas. Therefore, there appeared the necessity for the development of nonperturbing experimental methods of investigations for studying the structures and the local inhomogeneities in nonneutral plasmas. These methods were developed depending on the specific type of nonneutral plasma and on the geometry of containing device. The experimental methods of investigations described in [1] were developed for studying the local inhomogeneities in gas-discharge nonneutral electron plasma. The use of these methods allowed to discover the solitary vortex structures in such plasma, and to study their properties, behavior and the processes initiated by them. The present paper gives the results of experimental investigations of the processes of formation, evolution and dynamics of vortex structures. In Sec. II the parameters of vortex structures are given and the periodic sequence of evolution of vortex structure is described for different geometries of discharge device. In Sec. III the process of formation of the solitary vortex structure in the geometry of inverted magnetron is studies. In Sec. IV the dynamics of vortex structure in the magnetron geometry of discharge device is investigated. In Sec. V the mechanism of formation of vortex structure is discussed and the comparison is made between the processes of vortex structure evolution in gas-discharge nononeutral electron plasma and the results of simulation.

## II. Solitary vortex structures

The experiments were carried out in three geometries of discharge device: in magnetron, in inverted magnetron and in the Penning cell (Fig.1). The radius of external electrode was 3.2 and 3.7 cm, the radius of internal electrode – 0.9, 1.0 and 2.0 cm, the length of electrodes was 7.0 cm. The parameters of the discharge at the investigation of vortex structures varied within the following limits: the magnetic field $B = 1-2 kG$, the discharge voltage $V = 1-4 kV$, the pressure of neutral gas (argon) $p = 1 \times 10^{-6} - 2 \times 10^{-4} Torr$. The inhomogeneity of the magnetic field on the anode length was $\Delta B / B = 0.002$.

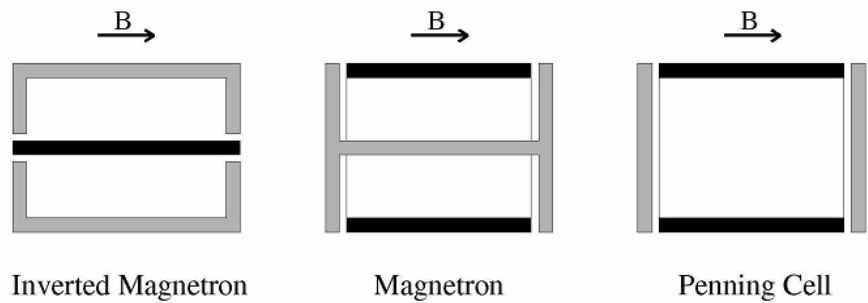

Fig.1. Geometries of discharge device. (black – anode)

To investigate the inhomogeneities the method of two wall probes was used [2], consisting in simultaneous measurement of signals from the wall probes of the anode and of the cathode during the rotation of vortex structure about the axis of discharge device. This method allows to follow continuously the trajectory and the charge of one or several vortex structures during a long period of time. In combination with the measurement of ejection of electrons from the vortex structure [3,4], the method of two wall probes gives the possibility to determine the parameters of vortex structures, to investigate their formation, interaction and dynamics. The measuring technique is described in more detail in [1]. The experimental results shown below are mostly obtained in the geometries of magnetron and inverted magnetron. In these geometries, first, it is convenient to use the method of two wall probes and, second, the electron ejection from plasma to the end cathodes is not "mixed" with the ion current. As for the Penning cell, in the cases when the density of ions can be neglected and when the thickness of electron sheath is less than the anode radius, the behavior of vortex structures in the Penning cell appeared to be similar to the behavior of vortex structures in the magnetron. Generally, it should be noted that the characteristics of vortex structures, as well as the processes of their formation, interaction and dynamics have lots in common for all three geometries of discharge device. However, depending on the geometry and on the pressure any of these processes becomes predominant creating the favorable conditions for its investigation.

The results of experimental investigations showed that in the sheath of gas-discharge nonneutral electron plasma there always presents the inhomogeneities consisting in dense compact structures stretched along the magnetic field [2-7], the lifetime of which is much greater than the time of electron-neutral collisions. Figure 2 (left) shows the oscillograms of oscillations of electric field on the wall probes of the anode (upper) and of the cathode (lower) at the rotation of one stable structure about the axis of the discharge device. The right side of the same figure shows the oscillograms of oscillations of electric field on the anode wall probe (upper) and of the current of electron ejection through the radial slit in the end cathode (lower). The slit is located at the same azimuth as the wall probes.

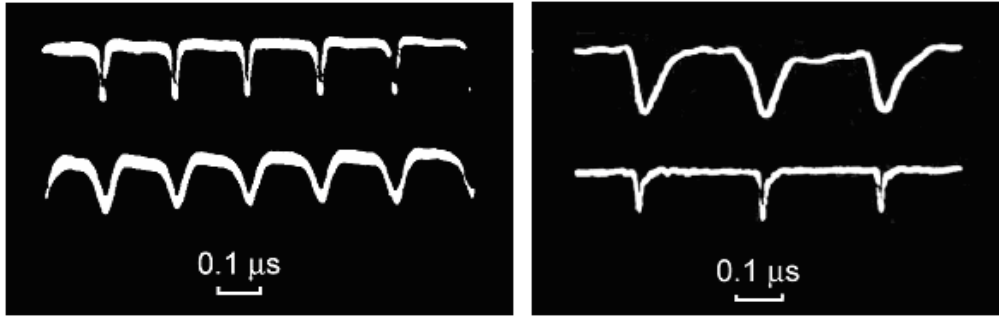

Fig.2. Oscillograms of the signals from the wall probes and from the radial slit
Left: $r_a = 3.2 cm$; $r_c = 1.0 cm$; $L = 7 cm$; $B = 1.5 kG$; $V = 1.0 kV$; $p = 2 \times 10^{-5} Torr$
Right: $r_a = 3.2 cm$; $L = 7 cm$; $B = 1.9 kG$; $V = 1.0 kV$; $p = 2 \times 10^{-5} Torr$

As it follows from the experimental measurements, the average radius of inhomogeneities is equal to $(0.2 - 0.3) cm$ [2-5], being much less than the dimensions of discharge gap ($\sim 2 cm$). The average density of electrons in inhomogeneities is significantly higher than the average density of sheath electrons. Hence, each inhomogeneity has its own electric field and, as a result, rotates about its own axis alongside with the rotation about the axis of discharge device. According to the estimations, the angular velocities of rotation of inhomogeneities about their own axes are much higher than the angular velocities of their rotation about the axis of discharge device. If in the sheath there are several inhomogeneities, they behave independent to each other: they move on different orbits with different angular velocities [2]. All this, allows to consider the inhomogeneities of gas-discharge nonneutral electron plasma as the solitary vortex structures.

The investigation of the processes of formation, evolution and dynamics of vortex structures showed that the behavior of these structures strongly depends on the pressure of neutral gas and on the type of geometry of discharge device. At pressures $p < 10^{-5} Torr$, both in magnetron and in inverted magnetron there is one quasi-stable vortex structure for a large period of time. However, the behavior and evolution of such structure are different for each geometry of discharge device. Figure 3 shows the oscillograms of oscillations of electric field on the anode wall probe (upper) and of the current of electron ejection to the end cathodes (lower) in magnetron (left) and in inverted magnetron (right).

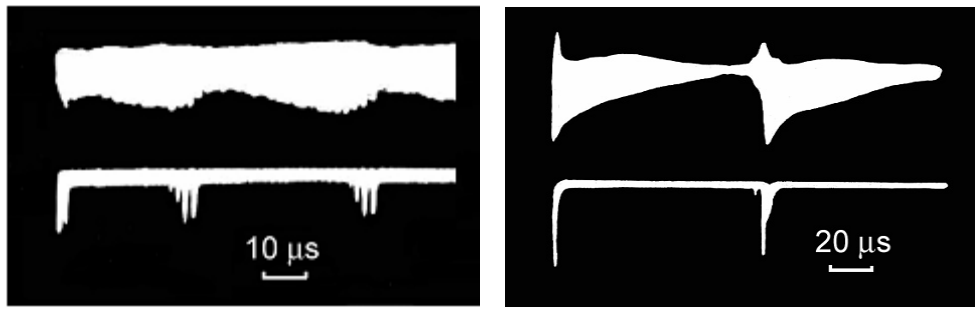

Fig.3 Periodical evolution of vortex structures in magnetron and in inverted magnetron
Left: $r_a = 3.2 cm$; $r_c = 1.0 cm$; $L = 7 cm$; $B = 1.2 kG$; $V = 1.5 kV$; $p = 6 \times 10^{-6} Torr$.
Right: $r_a = 1.0 cm$; $r_c = 3.2 cm$; $L = 7 cm$; $B = 1.8 kG$; $V = 0.9 kV$; $p = 1 \times 10^{-5} Torr$

As it follows from the oscillograms, in both geometries the periodically repeated processes take place with the period of repetition of the order of time of several tens of electron-neutral collisions.

In the magnetron the amplitude of electric field oscillations on the anode wall probe increases slowly. This can be caused both, by the increase of the charge of vortex structure and by its slow approach to the anode surface. At the definite moment of time, there appears an instability accompanied by the pulse ejections of electrons from the vortex structure to the end cathodes, as a result of which the vortex structure losses a part of its charge. Then, the charge of vortex structure starts again to increase until the following instability appears [2].

In the inverted magnetron, on the contrary, first there appears a diocotron instability, as a result of which the vortex structure is formed. The process of vortex structure formation is accompanied by an abrupt increase of the amplitude of electric field oscillations on the anode wall probe and by the pulse ejection of electrons to the end cathodes. Then, the amplitude of electric field oscillations decreases slowly, giving the evidence of a slow decrease of the charge of vortex structure. At the same time, the density of electron sheath increases until it reaches the "critical" value, at which there appears the next diocotron instability [4,6].

Below, by using the results of experimental investigations let us consider in more detail the processes of formation, evolution and dynamics of vortex structures in gas-discharge nonneutral electron plasma. The mechanism of electron ejection from the vortex structures to the end cathodes will be considered in the next paper.

**III. Formation of vortex structure**

The formation of vortex structures is convenient to observe in the geometry of inverted magnetron. In Fig.4 (left) the oscillograms of electric field oscillations on the wall probes of the anode (upper) and of the cathode (lower) are given showing a full cycle of appearance and evolution of the vortex structure in the inverted magnetron. Oscillograms on the right show how the ion current (upper) and the current of electron ejection to the end cathodes are changed. The process of evolution of the electron sheath proceeds in the following way: periodically, in the sheath there appears the diocotron instability leading to the formation of quasi-stable vortex structure. The formation of quasi-stable vortex structure is accompanied with the ejection of electrons from the electron sheath to the end cathodes. Then, the vortex structure" decays" slowly, and the density of electron sheath increases, this is evidenced by the increase of the density of ion current. When the density of electron sheath reaches the "critical" value, the diocotron instability appears again.

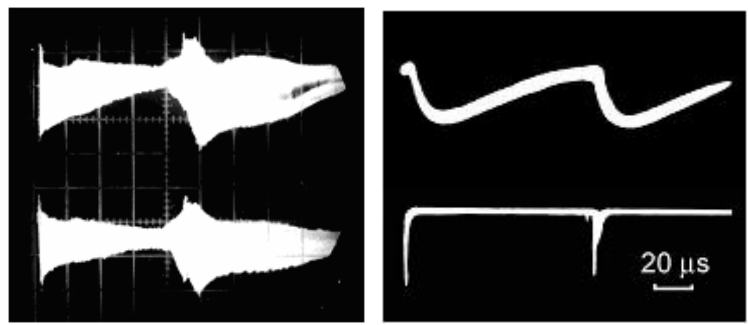

Fig.4 Periodical processes in the inverted magnetron
$r_a = 0.9 cm$; $r_c = 4 cm$; $L = 5 cm$; $V = 5 kV$; $B = 1 kG$; $p = 8 \times 10^{-5} Torr$.

Let us consider in detail the process of formation of quasi-stable vortex structure. In Figs. 5 and 6 the oscillograms of the fragments of this process are given. The upper oscillograms show the oscillations of electric field on the anode wall probe, and the lower oscillograms – on the cathode wall probe. Figure 5 shows the process of origination of vortex structure from the diocotron instability. The comparison of the amplitudes of oscillations on the anode and on the cathode give evidence of the fact that the whole process proceeds on the one and the same drift orbit, and the radial oscillations of the vortex structures are absent. On the initial parts of the oscillograms the strongly nonlinear oscillations of diocotron instability at the mode $l = 1$ are seen. Then, the oscillations increase abruptly and in the perturbed region of electron sheath a hole is formed. This is evidenced by the increase and by the narrowing of the positive half-period of oscillations and by the simultaneous widening of the negative half-period. However, very soon, just during several rotations of the inhomogeneity about the anode the pattern is changed cardinally. The hole is widened azimuthally, or strictly speaking, the electrons are bunched at the point being diametrically opposite azimuthally and form a clump. This is evidenced by the increase and by the narrowing of the negative half-period and, correspondingly, by the decrease and by the widening of the positive half-period of oscillations.

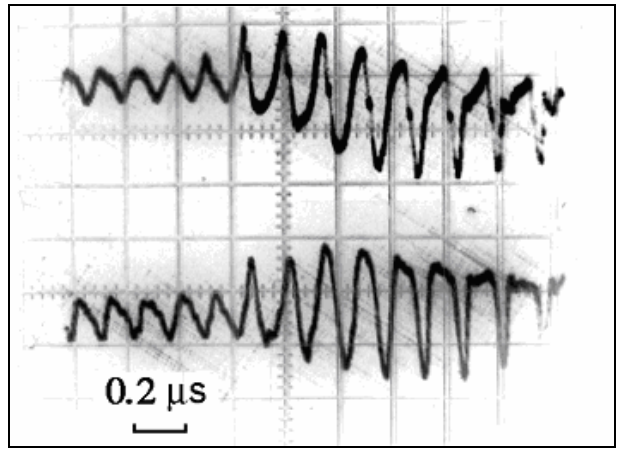

Fig.5. Process of formation of vortex structure
$r_a = 2.0 cm$; $r_c = 3.2 cm$; $L = 7 cm$; $B = 1.5 kG$; $V = 1.0 kV$; $p = 2 \times 10^{-5} Torr$

However, the formation of solitary vortex structure is not finished yet. The vortex structure has a tail as is evidenced by the slope and the ripple of the widened positive half-period of oscillations on the final part of the oscillograms in Fig.5. Figure 6 shows the further process of formation of vortex structure. The ripple of the tail is increased and a lot of small irregular inhomogenities are formed. These inhomogeneities approach each other, merge and form one more clump being less than the main one and moving with the less angular velocity (Fig.6, left).

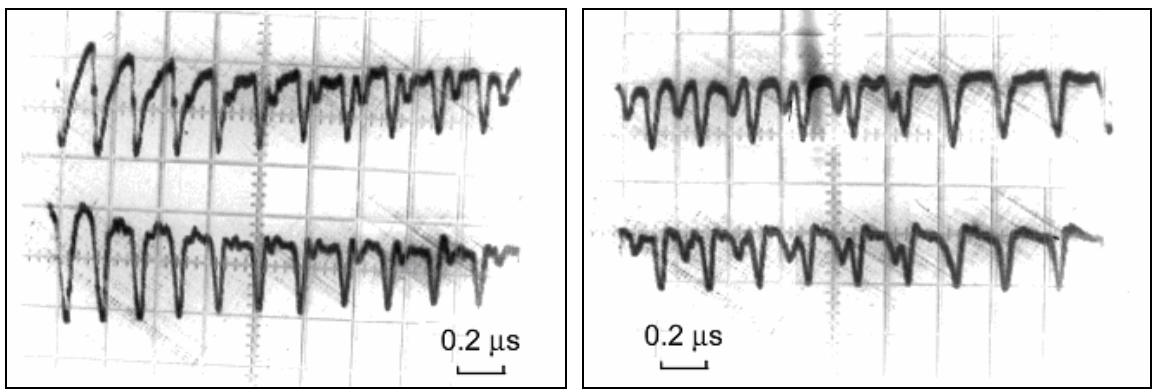

Fig.6. Formation of quasi-stable vortex structure
$r_a = 2.0 cm$; $r_c = 3.2 cm$; $L = 7 cm$; $B = 1.5 kG$; $V = 1.0 kV$; $p = 2 \times 10^{-5} Torr$

The main clump overtakes the second clump and absorbs it (Fig.6, right). At this point, the formation of the single stable vortex structure is finished. The whole process of formation of the stable vortex structure beginning with the formation of a hole up to the merging of two clumps takes place for the period of time much less than the time of electron-neutral collisions and is accompanied with the ejection of electrons from the electron sheath to the end cathodes. Thus, at the formation of vortex structure the decrease of the charge and, correspondingly, of the density of electron sheath takes place.

After the stable (more exactly, the quasi-stable) vortex structure is formed, the process of its slow "decay" is followed proceeding for the period of time much more than the time of electron-neutral collisions. Here, one should pay attention to the surprising fact: the "decay" of vortex structure despite the electron-neutral collisions is not accompanied by its diffusion. Figure 7 shows the fragments of electric field oscillations on the anode wall probe for two moments of time: when the stable vortex structure is fully formed (left oscillogram) and when it is near to the full decay (right oscillogram).

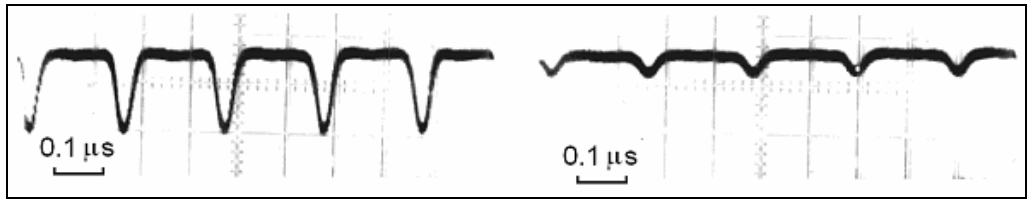

Fig.7. Decay of vortex structure
$r_a = 2.0 cm$; $r_c = 3.2 cm$; $L = 7 cm$; $B = 1.5 kG$; $V = 1.0 kV$; $p = 2 \times 10^{-5} Torr$

At low pressures ($\sim 10^{-6} Torr$) the vortex structure has the time to disappear fully and then, during some period of time (until the appearance of the next diocotron instability), the electron sheath remains unperturbed (Fig.8, left). Here, as in the Fig.3, the upper oscillogram presents the oscillations of the electric field on the anode wall probe, and the lower one – the current of electron ejection to the end cathodes.

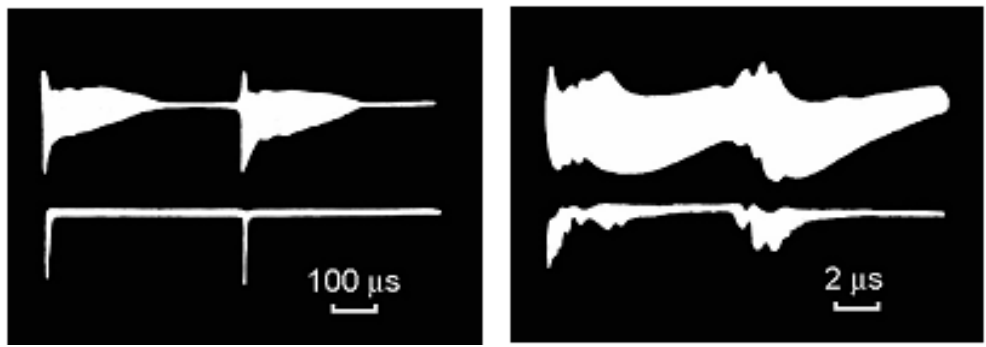

Fig.8. Periodic evolution of vortex structures in inverted magnetron
$r_a = 1.0 cm$; $r_c = 3.2 cm$; $L = 7 cm$; $B = 1.8 kG$; $V = 0.9 kV$; $p = 2 \times 10^{-6}, 1 \times 10^{-4} Torr$

However, at pressures of the order or more than $5 \times 10^{-5} Torr$ in the inverted magnetron there exist simultaneously several vortex structures, and more is the pressure the more is their number. Probably, this is caused by the fact that the velocity of "decay" of vortex structure increases with the pressure more slowly than the velocity of the increase of electron sheath density. Hence, a new vortex structure formed as a result of the next diocotron instability appears earlier than the preceding one has time to decay (Fig.8, right). Under such conditions, in the electron plasma the interaction of vortex structures becomes the dominant process.

**IV. Dynamics and evolution of vortex structure**

Now, let us study in detail the dynamics of vortex structure in the magnetron geometry of discharge device. At pressures $p < 10^{-5} Torr$, in the magnetron there exists only one vortex structure. However, its drift orbit periodically becomes instable [2]. At the definite moments of time the small radial oscillations of the vortex structure increases sharply. Figure 9 presents the oscillograms of electric field oscillations on the anode (upper) and cathode (lower) wall probes during the "orbital" instability.

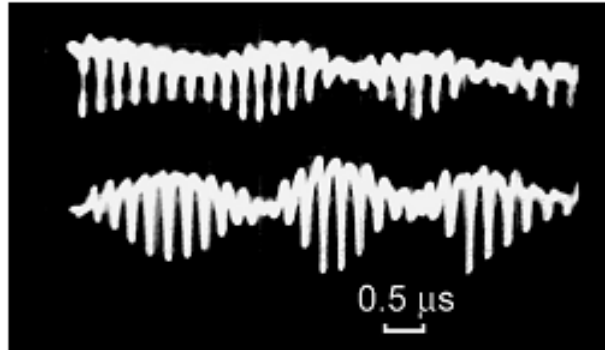

Fig.9. Orbital instability in magnetron
$r_a = 3.2 cm\,;\, r_c = 1.0 cm\,;\, L = 7 cm\,;\, B = 1.2 kG\,;\, V = 1.5 kV\,;\, p = 5\times 10^{-6} Torr$.

Each negative pulse on the oscillograms corresponds to the passing of vortex structures by the wall probes. As it is seen from the figure, the period of radial oscillations of the vortex structure is much more than the period of its rotations about the axis of the discharge device. Therefore, the vortex structure moves in a spiral, approaching periodically the anode or the cathode. The amplitude of radial oscillations of the vortex structure is sufficiently large and, sometimes, reaches 1 cm. Figure 10 shows the processes of development (left oscillogram) and decay (right oscillogram) of the "orbital" instability. Here, like Fig.9, the oscillograms of oscillations of electric fields on anode (upper) and cathode (lower) wall probes are shown.

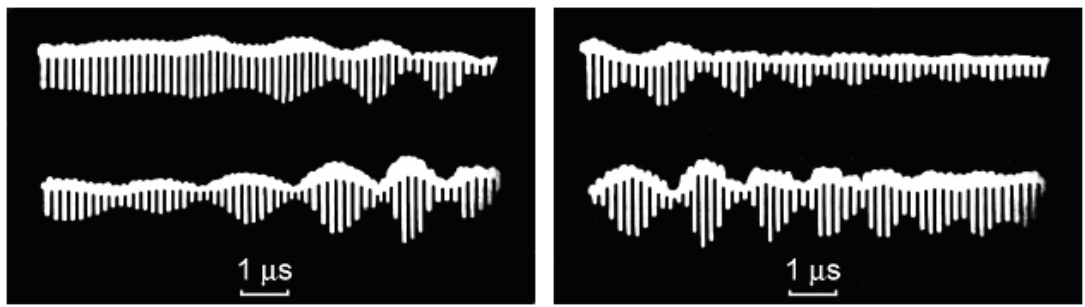

Fig.10. The motion of vortex structure during the orbital instability
$r_a = 3.2 cm\,;\, r_c = 1.0 cm\,;\, L = 7 cm\,;\, B = 1.5 kG\,;\, V = 3 kV\,;\, p = 5\times 10^{-6} Torr$

During the orbital instability, the charge of vortex structure decreases because each approach of vortex structure to the cylindrical cathode is accompanied by the ejection of electrons from the vortex structure to the end cathodes, and the duration of instability is less than the time of electron-neutral collisions. Completing 5-8 radial oscillations and losing about one third of its charge, the vortex structure "calms down". The orbit of vortex structure stabilizes. However, now, the radius of vortex structure orbit becomes by 5-6 mm less than it was before starting the instability. The orbit displacement towards the cathode during the orbital instability is well seen in Fig.10, if we compare the relation of oscillation amplitudes on wall probes of the anode and the cathode before and after the orbital instability. In Fig.11 the approximate variation of vortex structure orbit during the orbital instability is shown, on the left for magnetron [2] and on the right for Penning cell [7].

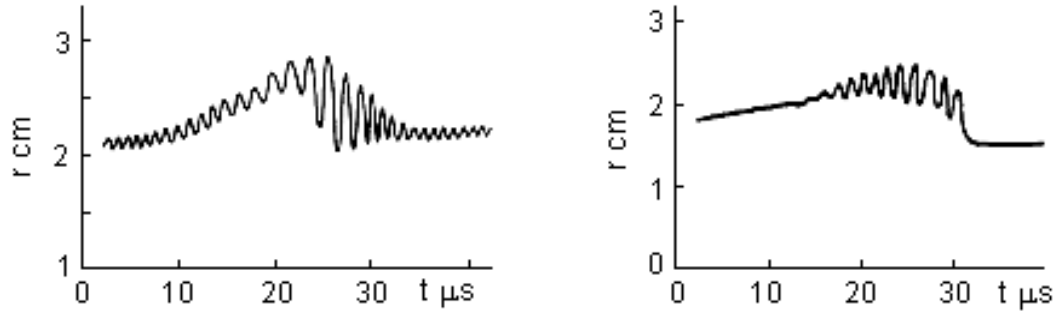

Fig.11. Radial oscillations of vortex structure in magnetron and in Penning cell [2,7]
Left: $r_a = 3.2 cm$; $r_c = 1.0 cm$; $L = 7 cm$; $B = 1.5 kG$; $V = 3.0 kV$; $p = 5\times 10^{-6} Torr$
Right: $r_a = 3.2 cm$; $L = 7 cm$; $B = 1.5 kG$; $V = 2.0 kV$; $p = 4\times 10^{-6} Torr$

After completion of orbital instability, the charge of vortex structure and the radius of its drift orbit begin to increase slowly. This lasts during the time much more than the time of electron-neutral collisions and continues until the orbital instability repeats again.

The orbital instability in Penning cell takes place in the same way as in magnetron geometry, though, there are several differences. The number of radial oscillations of vortex structure in Penning cell is more than in magnetron geometry (of the order of ten and more). The decay of radial oscillations in Penning cell in the most cases proceeds in the same way as in magnetron geometry, namely, the amplitude of radial oscillations of vortex structure and the amplitude of ejection of electrons decrease gradually (Fig.12, left). Here, the upper oscillogram shows the oscillations of electric field on the anode wall probe, and the lower one – the current of electrons to the end cathodes.

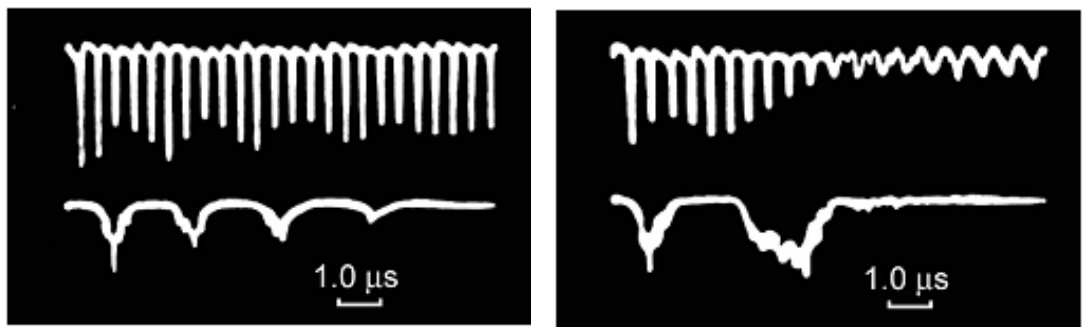

Fig.12. Orbital instability in Penning cell
$r_a = 3.2 cm$; $L = 7 cm$; $B = 2.0 kG$; $V = 2.0 kV$; $p = 6\times 10^{-6} Torr$

However, in some cases, the last ejection of electrons appears to be the greatest, and the amplitude of electric field oscillations on the anode wall probe decreases strongly (Fig.12, right). This indicates that sometimes the vortex structure comes quite close to the axis of Penning cell causing thus the great losses of electrons of vortex structure.

At pressures of the order or more than $5\times10^{-5} Torr$ in magnetron as well as in inverted magnetron the several vortex structures exist simultaneously and more is the pressure the more is their number. The vortex structures move on different orbits with different angular velocities and, therefore, periodically approach each other [2]. Figure 13 shows the oscillograms of electric field oscillations on the wall probes of the anode (upper) and of the cathode (lower) in the case of motion of two vortex structures about the axis of discharge device. According to the relation of amplitudes in the left figure, we can conclude that the structures move on different orbits and that the structure located near to the anode has a greater charge. In the right figure we observe the process of merging the vortex structures. The structure with a greater charge overtakes the structure with less charge and absorbs it.

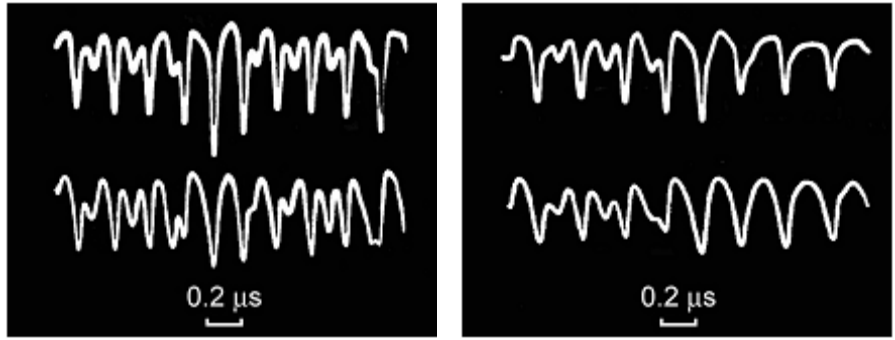

fig.13. Approach of vortex structures
$r_a = 3.2 cm$; $r_c = 1.0 cm$; $L = 7 cm$; $B = 1.5 kG$; $V = 1.5 kV$; $p = 1\times10^{-4} Torr$.

At approaching the vortex structures the pulse ejection of electrons takes place along the magnetic field both, from the vortex structures themselves, and from the electron background surrounding them. And this process of electron ejection becomes the prevailing process in gas-discharge nonneutral electron plasma at pressures of the order or more than $5\times10^{-5} Torr$. Nevertheless, alongside with the process of interaction of vortex structures, the processes of birth-and-death of vortex structures, as well as the process of radial oscillations of vortex structure are also kept. However, at such neutral gas pressures it is more difficult to observe these processes than at lower pressures.

**V. Discussion and conclusion**

The peculiarity of gas-discharge nonneutral electron plasma is the periodicity of the processes of evolution of vortex structures being especially pronounced at pressures of neutral gas $p<10^{-5} Torr$. Though, in the geometries of magnetron and inverted magnetron these processes proceed not similarly, they have the pronounced sequence: a short collisionless process connected with the instability and with the intense ejection of electrons to the end cathodes and a long evolution of quasi-stable vortex structure, during which the electron sheath (or the vortex structure) increases its own charge at the expense of ionization of neutral gas atoms by electrons.

The formation of vortex structure takes place as a result of diocotron instability appearing periodically in the geometry of inverted magnetron and non-regular in the magnetron geometry. This instability is well-known. It is investigated theoretically both, in Penning cell [8] and in geometries of magnetron and inverted magnetron [9]. As for the "orbital" instability, the mechanism of its origination is still not studied.

In gas-discharge electron plasma the annular electron sheath is formed naturally at the expense of electron-neutral collisions and ionization of neutral gas atoms by electrons. The density of electron sheath increases in time until it reaches the "critical" value at which there arises the diocotron instability. As it is shown in [9], at first, the diocotron mode $l=1$ should be excited, corresponding the minimum "critical" density of electron sheath. As a result, one quasi-stable vortex structure is formed just being observed experimentally. The formation of vortex structure is accompanied with pulse ejection of electrons to the end cathodes. This leads to the decrease of the average density of electron sheath and, thus, to the disturbance of the conditions of excitation of diocotron instability. Besides, the continuous ejection of electrons takes place to the end cathodes from the vortex structure and the adjacent regions of electron sheath. Therefore, until the vortex structure continues to exist, the density of electron sheath is limited to its minimum "critical" value. Such assumption has the experimental justification and therefore, it was used for developing the model of gas discharge electron sheath [10]. This model allowed to describe not only qualitatively, but also, to a considerable extent, quantitatively, the characteristics of electron sheath and the current characteristics of discharge both, in magnetron geometry and in the geometry of inverted magnetron.

In pure electron plasma the annular electron sheath is created artificially. Therefore, the diocotron instability can arise at any mode depending on the parameters of electron sheath. Correspondingly, several vortex structures are formed. For choosing the mode, usually the controlled initial disturbance of electron plasma is used. In the case of $l=1$ mode, the formation of vortex structure [11] takes place approximately in the same way as in gas discharge electron plasma. However, the following evolution of vortex structure in pure electron plasma strongly differs from that taking place in gas-discharge electron plasma.

Generally speaking, the equations describing the diocotron instability give the conditions of its appearance at one or another mode. The further process of appearance of turbulence state or of formation and evolution of vortex structure are generally studied by means of numerical simulation. However, so far, the simulation has been limited only to pure electron plasma. The gas-discharge electron plasma has its own peculiarities: ionization, self-sustaining regime, unlimited time of existence and, finally, the partial presence of ions. In the paper published recently [12], the simulation was made for the low-pressure discharge in magnetron geometry at about the same geometrical and electrical parameters of the discharge as in the experiments described above. At the simulation, the electron-neutral collisions, ionization, electron loss on the end cathodes (if their longitudinal velocity allowed to overcome the potential barrier), the secondary ion-electron emission from the cathode were considered, i.e. everything that takes place in the real discharge. The obtained results differ somehow from that observed experimentally. However, in general terms, they demonstrate quite well the main processes and the consequence of events. First, the electron sheath located near the anode is instable and is transformed into pronounced vortex structures. Second, the periodical consequence of processes takes place from the "rapid" formation of vortex structures to the "slow" transition to the chaotic distribution of electron density, from which the vortex structures are originated again. Third, the approach to each other, the merging and the approach of vortex structures to the cathode are accompanied with the pulses of electron current on the end cathodes. On the other hand, the transformation of chaotic distribution of electron density into symmetrically located pronounced vortex structures and their following stepwise relaxation are not in agreement with the experimental results in gas-discharge electron plasma and resemble the process of formation of "vortex crystals" in pure electron plasma [13].

Finally, the experiments described above, were carried out in three geometries of discharge device: magnetron, inverted magnetron and Penning cell. For the discharge parameters, when the thickness of electron sheath was less than the size of discharge gap, the vortex structures were always observed. Moreover, the vortex structures were always accompanied by the electron ejection to the end cathodes. Hence, the electron current on the end cathodes can serve as an indirect proof of the existence of vortex structures. From this it follows that the electron current to the cathode being observed practically by all investigators of the low-pressure discharge in crossed electric and magnetic fields can serve as an evidence of the existence of vortex structures. The existence of vortex structures in most devices with gas-discharge nonneutral electron plasma gives the evidence of the fact that they are the essential components of such plasma.